\documentclass[a4paper,11pt]{article}
\pdfoutput=1 

\usepackage{jcappub} 

\usepackage[T1]{fontenc} 

\title{\boldmath 
New method to revisit the gravitational lensing analysis of the Bullet Cluster using radio waves
}

\author[a,b]{Youngsub Yoon}
\author[a,b]{Jong-Chul Park}
\author[c,d]{Ho Seong Hwang}

\affiliation[a]{Department of Physics and Institute of Quantum Systems (IQS) \\Chungnam National University, Daejeon 34134, Republic of Korea}
\affiliation[b]{Particle Theory and Cosmology Group, Center for Theoretical Physics of the Universe, Institute for Basic Science (IBS),  Daejeon, 34126, Republic of Korea}
\affiliation[c]{Astronomy Program, Department of Physics and Astronomy,\\Seoul National University, Seoul 08826, Republic of Korea}
\affiliation[d]{SNU Astronomy Research Center,\\Seoul National University, Seoul 08826, Republic of Korea}

\emailAdd{youngsuby@gmail.com}
\emailAdd{jcpark@cnu.ac.kr}
\emailAdd{hhwang@astro.snu.ac.kr}

\abstract{
Gravitational lensing studies of the Bullet Cluster suggested convincingly in favor of the existence of dark matter. 
However, it was performed without the knowledge of the original orientation of each galaxy before gravitational lensing. 
A potential improvement to this issue lies in the measurement of the original orientation from the polarization direction of radio waves emitted from each galaxy. 
In this context, Francfort et al. derived a formula that can utilize the information about the original orientation of each galaxy to obtain what is called {\it shear}. 
However, we demonstrate that shear in their formula should be replaced by {\it reduced shear} when the change in sizes of images of galaxies is taken into account. 
As the previous gravitational lensing analysis of the Bullet Cluster used reduced shear, we suggest applying our improved formula directly for the reanalysis once we obtain the polarization direction of radio waves. 
In particular, we show that our new formula can yield a more accurate analysis than the previous one, if the polarization direction can be measured more precisely than $10^\circ$. 
Moreover, the approach discussed in this work is generically applicable to the gravitational lensing analysis of clusters, not only limited to the Bullet Cluster.
}

\begin{document}

\maketitle
\flushbottom

\section{Introduction}\label{introduction}

There are many observational results that favor the existence of dark matter.
One of the most convincing results is the gravitational lensing analysis of the Bullet Cluster~\cite{Clowe:2006eq}; matter present in the Bullet Cluster, be it baryonic matter or dark matter, distorts the images of galaxies behind the Bullet Cluster, by its gravitation. 
The authors of Ref.~\cite{Clowe:2006eq} analyzed such images to reconstruct the mass distribution at the Bullet Cluster, which did not coincide with the baryonic matter distribution obtained by X-ray image. 
Thus, they concluded that dark matter is responsible for the discrepancy.

In order to analyze the gravitational lensing effect, certain assumptions about the original images are necessary since the observed images of galaxies alone cannot determine the distortion.
In Ref.~\cite{Clowe:2006eq}, it is assumed that the average orientation of the original galactic images in each small patch of sky, where variables related to gravitational lensing are determined, is zero.
However, this can lead to errors if there are not enough galaxies in each patch. 
Although this represents the optimal approach based on the currently available observational data, the analysis requires a sufficient number of galaxies to statistically determine the gravitational lensing effect in each patch.
Otherwise, accidental skewing of the original galactic orientations could lead to skewed results.

However, it is now possible to determine the original orientation of galaxies from the polarization of the radio waves from each galaxy. 
The radio emission from each galaxy is known to have a polarization that is perpendicular to the major axis of its ellipticity~\cite{Stil:2008ew,Brown:2010rr}. 
While the orientation of a galactic image is rotated by gravitation, polarization is not. 
Therefore, even if the average original galactic orientations were distorted in a certain direction by accident, possibly due to the small number of galaxies in each small patch of sky, it would not bias the data, as long as we know the original orientation and therefore are able to compensate it. 
Thus, we can use this information to our advantage to measure the lensing effect more accurately, as pointed out in Refs.~\cite{Brown:2010rr, Francfort:2021oog, Brown:2011db}. 
Therefore, the position of dark matter at the Bullet Cluster may be corrected if we reanalyze the gravitational lensing effect with the help of the polarization data of radio waves, which would be available in the future~\cite{SKA:2018ckk, Brown:2015ucq, SKAMagnetismScienceWorkingGroup:2020xim}.

Regarding gravitational lensing analysis, Francfort et al. obtained a formula that can be employed in such a situation, i.e., in cases when the original orientations of galaxies are known~\cite{Francfort:2021oog}. 
However, their formula is for {\it shear}, a variable used in the gravitational lensing effect. 
This presents a hurdle in reanalyzing the gravitational lensing effect at the Bullet Cluster because the previous analysis used a variable called {\it reduced shear}. 
Therefore, we cannot directly use the formula given in Ref.~\cite{Francfort:2021oog} to repeat the analysis. 

Nevertheless, the formula obtained in Ref.~\cite{Francfort:2021oog} pertains to the reduced shear, rather than the shear itself:
in their derivation, they ignored the size change of galactic images in the gravitational lensing to simplify the calculation, and only considered the shape change. 
Here, we show that the shear in their formula should be replaced by the reduced shear if we consider the size change. 
This is a great advantage because this new formula for the reduced shear can be directly applied for the reanalysis of the gravitational lensing effect at the Bullet Cluster.

The organization of this paper is as follows. 
In Section~\ref{gravitationallensing}, we give a brief introduction to the gravitational lensing analysis, as a background to understand subsequent sections. 
In Section~\ref{currentmethod}, we explain the current method to estimate the gravitational lensing effect, in particular, the reduced shear.
In Section~\ref{newmethod}, we present our new method to estimate the reduced shear.
In Section~\ref{francfortsformula}, we review the main results of Ref.~\cite{Francfort:2021oog}, which illustrate how we can improve the gravitational lensing effect estimation using the polarization data of radio emissions. 
In particular, we will present a new formula for shear, derived in Ref.~\cite{Francfort:2021oog}. 
In Section~\ref{ournewformula}, we demonstrate that the formula for shear should be a formula for reduced shear. 
In addition, we recast the formula in terms of the second brightness moments already measured and used in the gravitational lensing analyses of the Bullet Cluster. 
In Section~\ref{discussion}, we conclude our paper with discussions.
In Appendix~\ref{appendixkeyformula}, we present a detailed derivation of our key formula for the reduced shear.
In Appendix~\ref{B}, we provide a comprehensive calculation of the error of the reduced shear obtained from our new formula.

\section{Key variables in gravitational lensing analysis}\label{gravitationallensing}

To understand how gravitational lensing effect is analyzed, it is important to know how its key variables are defined. 
Let $\theta_i$ ($i=1, 2$) be two orthogonal coordinates on the sky that denote the observed position of the image of a galaxy. 
Let $\beta_i$ be the position of a galaxy image, if there was no gravitational lensing. 
Then, we define the Jacobian map $A_{ij}$ by 
\begin{equation}
	A_{ij}\equiv \frac{\partial \beta_i}{\partial \theta_j}\,.
\end{equation} 
When there is no gravitational lensing, this matrix simply reduces to the identity matrix. 
Then, the convergence $\kappa$ and the shear $\gamma_1, \gamma_2$ are defined by the deviations of $A_{ij}$ from the identity matrix as follows,
\begin{equation}
	A=\left(\begin{array}{cc}
		1-\kappa +\gamma_1 & \gamma_2\\
		\gamma_2 & 1-\kappa-\gamma_1
	\end{array}\right)\,.
 \label{Aoriginal}
\end{equation} 
In other words, in the lowest order, the convergence $\kappa$ (also called {\it expansion}) denotes the shape-independent overall change in the size of the image of a galaxy, and the shear $\gamma$ denotes the shape change. 
Here, we see that the above matrix is symmetric. 
In general, the Jacobian matrix is not symmetric, which is denoted by an angle $\psi$ (called {\it twist})~\cite{Perlick:2004tq}. 
However, it is negligible as it is zero in the Schwarzschild  case~\cite{Francfort:2021oog}. 
It is also easy to see that only the convergence, not the shear, concerns the size change at first order. 
If we let $\gamma=\sqrt{\gamma_1^2+\gamma_2^2}$, the two eigenvalues of $A$ are given by 
\begin{equation}
	a_+=1-\kappa+\gamma,\quad a_-=1-\kappa-\gamma\label{a+a-}\,.
\end{equation}
Thus, the areal size of the original image is $a_+a_-=1-2\kappa+\mathcal O(\kappa^2)+\mathcal O(\gamma^2)$ times the one of the observed images.

Note that there is no direct way to determine the change of the size of image by observation, while there are ways to determine the shape change. 
Unlike the shape, no assumptions can be made about the size of the original (source) image. 
Therefore, a new variable should be introduced as follows: the reduced shear can be defined by
\begin{equation}
	g_\alpha\equiv\frac{\gamma_\alpha}{1-\kappa}
\end{equation}
for $\alpha=1,2$. 
Then Eq.~(\ref{Aoriginal}) can be rewritten as
\begin{equation}
	A=(1-\kappa)\left(\begin{array}{cc}
		1+g_1 & g_2\\
		g_2 & 1-g_1
	\end{array}\right)\,.\label{Ag}
\end{equation} 
As we cannot directly determine the overall factor $(1-\kappa)$ from the image, the only thing we can measure from the shape change is not the shear $\gamma_\alpha$ but the reduced shear $g_\alpha$.

After determining the reduced shear, we can obtain the convergence $\kappa$ through the following formula \cite{Kaiser:1994jm, Schneider:1994jj}:
\begin{equation}
	\nabla\ln(1-\kappa)=\frac{1}{1-g_1^2-g_2^2}	\left(\begin{array}{cc}
		1+g_1 & g_2\\
		g_2 & 1-g_1
	\end{array}\right)\left(\begin{array}{c}
		g_{1,1}+g_{2,2}\\g_{2,1}-g_{1,2}
	\end{array}\right)\,, \label{kappa}
\end{equation}
where $g_{a,b}$ denotes $\partial g_a / \partial \theta_b$. 
The right-hand side must be numerically integrated to find $\kappa$. 
According to the general relativity, $\kappa$ is not only convergence, but also what is called the {\it dimensionless surface mass density}. 
In other words, $\kappa$ is directly related to the mass distribution. 

This is how the authors of Ref.~\cite{Clowe:2006eq} analyzed the gravitational lensing effect of the Bullet Cluster to conclude that dark matter exists. 
In particular, the determination of the reduced shear $g_\alpha$ was necessary in their analysis. 
In the next section, we will focus on how they determined the reduced shear $g_\alpha$.

\section{Current method to estimate the reduced shear}\label{currentmethod}

\begin{figure}[ht]
	\centering
	\includegraphics[width=0.5\linewidth]{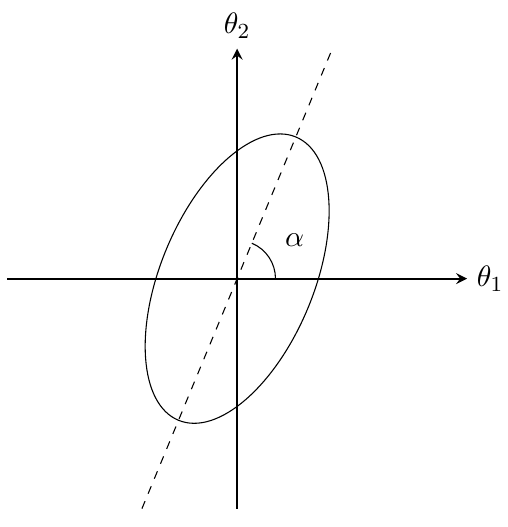}
	\caption{A schematic diagram to understand the relation between the orientation of a galaxy and its second brightness moments. 
    The ellipse depicts an image of galaxy.
    The origin of the coordinates is shifted to the center of the galaxy by $\bar\theta_1$ in the $\theta_1$ direction and $\bar\theta_2$ in the $\theta_2$ direction.
    Here, $Q_{22}$ is bigger than $Q_{11}$, and $Q_{12}$ is positive. }
    \label{phi}
\end{figure}

In the original gravitational lensing analysis at the Bullet Cluster~\cite{Clowe:2006eq}, the reduced shear was determined only from the observed ellipticity of galaxies which in turn can be obtained from the second brightness moments defined as follows~\cite{Blandford:1991edc, Bartelmann:1999yn},
\begin{equation}
	Q_{ij}=\frac{\int  I(\theta) (\theta_i-\bar \theta_i) (\theta_j-\bar \theta_j)d^2\theta}{\int I(\theta)d^2\theta}\,, \label{Qij}	
\end{equation}
where $I(\theta)$ is the brightness distribution. 
Here, $\bar\theta_i$ is the light-intensity-weighted center of the galaxy image, defined by
\begin{equation}
	\int I(\theta)(\theta_i-\bar\theta_i)d^2\theta=0\,.
\end{equation}
To understand the second brightness moments better, see Fig.~\ref{phi}.
In this case, $Q_{22}$ is bigger than $Q_{11}$, as the ellipse is more stretched along the $\theta_2$ direction than along the $\theta_1$ direction. 
In addition, $Q_{12}$ is positive because the major axis lies in the region $(\theta_1-\bar\theta_1)(\theta_2-\bar\theta_2)>0$, i.e., the first and the third quadrants.
In practice, because of noise and the presence of neighboring objects, $I(\theta)$ must be replaced by $I(\theta)w(\theta-\bar\theta)$ where $w$ is a weight function that quickly approaches zero, as $\theta$ moves away from $\bar\theta$. 

Then, the eigenvalues of the second brightness moments $Q_{ij}$ are given by $a^2$ and $b^2$ where $a$ is the semi-major axis and $b$ is the semi-minor axis. 
If we define new variables by the following relations
\begin{equation}
	Q_1\equiv Q_{11}-Q_{22},\quad Q_2\equiv 2 Q_{12},\quad T=Q_{11}+Q_{22}\,,\label{Q1Q2}
\end{equation}
the eigenvalues of our earlier matrix are given by
\begin{equation}
	a^2=\frac{T+\sqrt{Q_1^2+Q_2^2}}{2},\quad  b^2=\frac{T-\sqrt{Q_1^2+Q_2^2}}{2}\,. \label{a^2b^2}
\end{equation}
Then, from the following fact
\begin{equation}
	\frac{a^2-b^2}{a^2+b^2}=\frac{\sqrt{Q_1^2+Q_2^2}}{T}=\frac{|Q_1+iQ_2|}{T}\,,
 \label{Q1+iQ2T}
\end{equation}
we can be motivated to define the complex ellipticity as follows
\begin{equation}
	e_1+i e_2 \equiv \frac{Q_1+i Q_2}{T}\label{e1ie2}\,.
\end{equation}
Notice that this ellipticity has two real components, $e_1$ and $e_2$, which can be positive or negative. 
Thus, it is different from the usual ellipticity which can never be negative and is smaller than 1.
In Fig.~\ref{phi}, $\alpha$ is the angle between the $\theta_1$ axis and the major axis of the observed elliptical image of galaxy, which is given by 
\begin{equation}
	e_1+ie_2=\left(\frac{a^2-b^2}{a^2+b^2}\right) e^{2i\alpha}\,.
\end{equation}
Thus, we have a simple relation,
\begin{equation}
	\tan (2\alpha)=\frac{e_2}{e_1}=\frac{Q_2}{Q_1}\,. \label{tan2phi}
\end{equation}
For example, we find $Q_1=Q_{11}-Q_{22}<0$ and $Q_2=2Q_{12}>0$ in Fig.~\ref{phi}, which implies that $\tan (2\alpha)$ is negative, which agrees with the fact that $45^\circ<\alpha<90^\circ$.

Gravitational lensing changes $e_\alpha$ by the following amount
\begin{equation}
	\delta e_\alpha = P^\gamma_{\alpha\beta}	g_\beta\quad (\alpha, \beta = 1, 2)\,,
\end{equation}
where $P^\gamma_{\alpha \beta}$ defines the shear susceptibility tensor which one can calculate from each galaxy image, and $g_\beta$ is the reduced shear~\cite{Kaiser:1994jb}. 
Here, $\gamma$ denotes ``shear'' and is not an index.
If we denote the image of the source without gravitational lensing by the label $(s)$, the change in the complex ellipticity is given by $\delta e_\alpha = e_\alpha - e^{(s)}_\alpha$. 
Therefore, the reduced shear $g_\beta$ is given by
\begin{equation}
	g_\beta = (P^{\gamma}_{\alpha\beta})^{-1} (e_\alpha-e^{(s)}_\alpha)\,.\label{gamma}
\end{equation}
If the orientations of galaxies are random, $e_\alpha^{(s)}$ satisfies
\begin{equation}
	\langle e_\alpha^{(s)}\rangle=0\,.
\end{equation}
Therefore, assuming the full randomness $\langle (P^{\gamma}_{\alpha\beta})^{-1} e_\alpha^{(s)} \rangle=0$, we have
\begin{equation}
	g_\beta=\langle (P^{\gamma}_{\alpha\beta})^{-1} e_\alpha \rangle\,. \label{gbeta}
\end{equation}
However, this formula is not practical, as the shear susceptibility tensor is very noisy, biasing the data. 
Therefore, some additional complicated statistical procedures are necessary, such as averaging the shear susceptibility tensors of galaxies with a similar light profile which is supposed to have a similar shear susceptibility tensor and then calculating in turn a weighted average of the reduced shear obtained by such a method~\cite{Clowe:2005zg}.

In this section, we briefly reviewed how the authors of Ref.~\cite{Clowe:2006eq} obtained the reduced shear. 
In particular, we have seen that they had to assume that the average of the complex ellipticity is zero, as they did not have any information about the original orientations of galaxies before the gravitational lensing. 
While it is a reasonable assumption, it could be a problem because the average complex ellipticity could significantly deviate from zero if there are not enough galaxies. 
In particular, the error in one-dimensional (1D) reduced shear can be estimated as
\begin{equation}
	\sigma_g = \frac{\langle (P^{\gamma})^{-1} e_{\mathrm{rms}}\rangle}{\sqrt N} = \frac{g_{\mathrm{rms}}}{\sqrt N}\,,
\end{equation}
where $N$ is the number of galaxies in each patch. 
We will see later that the precise value for $g_{\mathrm{rms}}$ does not matter for the reduced shear error comparison between the current method and the new one. 
For the time being, we will plug in 0.27 for this value, as the rms 1D reduced shear for $z\sim 0.5$ is around 0.27 according to Table 2 in Ref.~\cite{Clowe:2005zg}. 
The value for the Bullet Cluster, which is in the redshift around 0.3, should not be drastically different. 
Therefore, we have
\begin{equation}
	\sigma_{g}=\frac{0.27}{\sqrt N}\,.
 \label{appendixa}
\end{equation}
In the following sections, we will elucidate how we can better estimate the reduced shear, if we have information about the original orientations of galaxies.

\section{New method to estimate the reduced shear}\label{newmethod}

\subsection{A formula for the shear with prior knowledge of the radio wave polarization}\label{francfortsformula}

\begin{figure}[ht]
	\centering
	\includegraphics[width=0.5\linewidth]{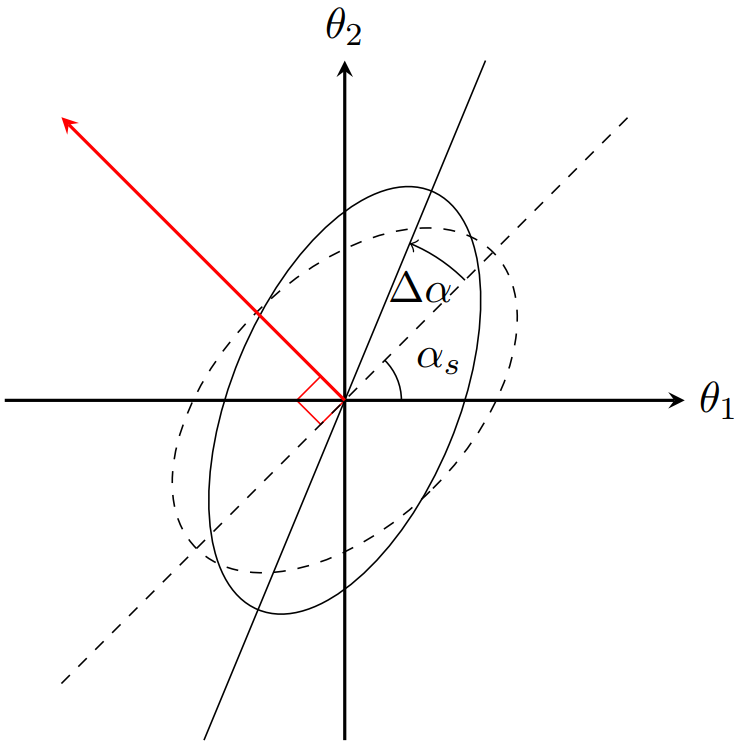}
	\caption{The perpendicularity between the polarization direction of the radio wave and the major axis of the original galactic image.
    The red arrow and the dotted straight line correspond to the polarization direction of the radio wave and the major axis of the original galactic image, respectively.
    The angle shift of the galactic image due to gravitational lensing is denoted by $\Delta\alpha$.}
 \label{rotation}
\end{figure}

In Ref.~\cite{Francfort:2021oog}, the authors derived a formula for the shear that can be used if the original orientation of a galaxy before gravitational lensing is known. 
The original orientation can be determined from the radio wave, which the electrons moving in the magnetic field of a galaxy emit by synchrotron radiation.
As the magnetic field in a galaxy is dominantly in the galactic plane, the polarization of the radio wave coming from each galaxy is approximately perpendicular to the major axis of its (un-lensed) image~\cite{Stil:2008ew, Brown:2010rr}, 
\begin{equation}
	\theta_{\mathrm{pol}}  \simeq  \alpha_s+90^\circ\,,
\end{equation}
where $\alpha_s$ is $\alpha$ in Eq.~(\ref{tan2phi}) if there is no gravitational lensing. 
In Fig.~\ref{rotation}, the red arrow denotes the polarization direction of the radio wave, and the dotted lines denote the original (i.e., un-lensed) galactic image and its major axis. 
In other words, if we know $\theta_{\mathrm{pol}}$, we also know $\alpha_s$. 
Note that the rotation of the polarization direction due to gravitational lensing is negligible, as it is $\mathcal O(\Delta\theta)$ effect where $\Delta\theta$ is the deflection angle~\cite{Hou}.
The deflection angle is $\lesssim 30''$ for clusters~\cite{Schneider}. 
In the figure, the solid ellipse and line represent the observed image of the galaxy and its major axis, respectively. 
From the observed image and the radio wave polarization direction, we can determine the angle shift, denoted by $\Delta \alpha\equiv \alpha_o-\alpha_s$ in the figure. 

The authors of Ref.~\cite{Francfort:2022laa} obtained the following relation between the shear $\gamma_\alpha$ and the angle shift $\Delta\alpha$,\footnote{In Ref.~\cite{Francfort:2022laa}, the numerator and the denominator on the right-hand side of the relation are reversed.
However, the correct one is Eq.~(\ref{gamma2gamm1}), which is clear from Eq.~(22) in Ref.~\cite{Francfort:2021oog}. 
The authors of these papers confirmed this mistake.}
\begin{equation}
	\gamma_2 \cos 2\alpha_s-\gamma_1 \sin 2\alpha_s=\frac{\varepsilon^2}{2-\varepsilon^2}\Delta \alpha\,, 
  \label{gamma2gamm1}
\end{equation}
where $\varepsilon$ is the usual ellipticity given by\footnote{Whether we use the ellipticity of the observed galactic image or the original image before gravitational lensing does not result in a significant difference, as the two ellipticities are approximately the same in weak lensing. 
However, as the former is directly observable, we use it in our equation.}
\begin{equation}
	\varepsilon=\sqrt{1-\frac{b^2}{a^2}}\,,
  \label{usualellipticity}
\end{equation}
where $a$ is the semi-major axis and $b$ is the semi-minor axis.
Note that the only unknowns in Eq. (\ref{gamma2gamm1}) are $\gamma_1$ and $\gamma_2$. 
Therefore, Eq. (\ref{gamma2gamm1}) is a variant of linear regression. 
In the next subsection, we will explain how Eq. (\ref{gamma2gamm1}) can be improved.

\subsection{New formula for the reduced shear}\label{ournewformula}

While Eq.~(\ref{gamma2gamm1}) is correct as long as the approximation used in the derivation is valid, it can lead to an error if the approximation is no longer valid. 
In fact, the image magnification is ignored in the analysis of Ref.~\cite{Francfort:2021oog} to make the calculation simpler. 
Thus, it is clear from the argument presented in Section~\ref{gravitationallensing} that the shear in the relation from Ref.~\cite{Francfort:2021oog, Francfort:2022laa} would be replaced by the reduced shear if the image magnification had not been ignored, i.e., the convergence, as the shape change is not given by the shear but by the reduced shear.

For a more concrete argument, we have to compare our earlier Jacobi map Eq.~(\ref{Aoriginal}), i.e., Eq.~(\ref{Ag}) with the following Jacobi map in Ref.~\cite{Francfort:2021oog}:\footnote{Here, we flip the sign in front of $\gamma_1$ to follow the sign convention of Eq.~(\ref{Aoriginal}) which is from Ref.~\cite{Clowe:2006eq}, where the existence of dark matter is strongly suggested based on the gravitational lensing analysis of the Bullet Cluster.}
\begin{equation}
	D=D_s\left(\begin{array}{cc}
		\cos\psi & -\sin\psi\\
		\sin\psi & \cos\psi
	\end{array}\right) \exp \left(\begin{array}{cc}
		\gamma_1 & \gamma_2\\
		\gamma_2 & -\gamma_1
	\end{array}\right)\,.\label{D}
\end{equation}
Unlike our earlier Jacobi map, it is defined by the matrix that relates the observed angle to the position on the Sachs screen at the source. 
Here, $D_s$ is a matrix that transforms the angular size into the distance on the Sachs screen, and $\psi$ quantifies the non-symmetricity of the Jacobi map. The exponential in the above equation is given at the leading order by 	
\begin{equation}
	\exp \left(\begin{array}{cc}
		\gamma_1 & \gamma_2\\
		\gamma_2 & -\gamma_1
	\end{array}\right)=\left(\begin{array}{cc}
		1+\gamma_1 & \gamma_2\\
		\gamma_2 & 1-\gamma_1
	\end{array}\right)\,.\label{expgamma}
\end{equation}
If we consider the two different definitions of the Jacobi matrix, Eq.~(\ref{Aoriginal}) and Eq.~(\ref{D}), and that $\psi$ is negligible in the lowest order, it is easy to see that the right-hand side of Eq.~(\ref{expgamma}) must be replaced by Eq.~(\ref{Aoriginal}), i.e., Eq.~(\ref{Ag}). 
Then, the magnification factor $(1-\kappa)$ can be combined with $D_s$, which relates the two different size scales, one at the position and another at the Sachs screen, and $\gamma_1$, $\gamma_2$ can be replaced by $g_1$, $g_2$. 
Moreover, the authors of Ref.~\cite{Francfort:2021oog} noted  $\gamma=\log(a_+/a_-)/2$, where $a_+$ and $a_-$ are the two eigenvalues of the Jacobi map, Eq.~(\ref{D}). 
Remembering that our Jacobi map is given by Eq.~(\ref{Aoriginal}), we see that $\log(a_+/a_-)/2$ is in fact given by
\begin{equation}
	\frac 12 \log\left(\frac{1-\kappa+\gamma}{1-\kappa-\gamma}\right)=\frac 12 \log\left(\frac{1+g}{1-g}\right)= g + \mathcal O(g^3)\,,
\end{equation}
where $g=\sqrt{g_1^2+g_2^2}$. 
Thus, we recover again that $\log(a_+/a_-)/2$ is not the shear but the reduced shear. 
The magnification factor $1-\kappa$ cancels out, which is also apparent from Eq.~(\ref{Ag}).

Meanwhile, the right-hand side of Eq.~(\ref{gamma2gamm1}) can be easily related to the variables introduced in Section~\ref{currentmethod}. 
From Eqs.~(\ref{tan2phi}) and (\ref{usualellipticity}), we have
\begin{equation}
	\frac{\varepsilon^2}{2-\varepsilon^2}=\frac{a^2-b^2}{a^2+b^2}=|e_1+ie_2|\,.
  \label{2-esquareesquare}
\end{equation}
Consequently, Eq.~(\ref{gamma2gamm1}) should be replaced by\footnote{There is an overall negative sign compared to Eq.~(\ref{gamma2gamm1}) as the sign convention for the shear and the reduced shear in Ref.~\cite{Francfort:2021oog} is different from ours.}
\begin{equation}
	g_2 \cos 2\alpha_s-g_1 \sin 2\alpha_s=-|e_1+ie_2|\Delta\alpha\,. 
  \label{g2g1}
\end{equation}
We provide a detailed derivation of this formula in Appendix ~\ref{appendixkeyformula} by closely following the derivation of Eq.~(22) in Ref.~\cite{Francfort:2021oog}.

Similarly as in Eq.~(\ref{gamma2gamm1}), the only unknowns in Eq.~(\ref{g2g1}) are $g_1$ and $g_2$ as all the other quantities are measurable. 
Therefore, Eq.~(\ref{g2g1}) is a variant of linear regression. Thus, we can easily obtain $g_1$ and $g_2$ by statistical analysis.\footnote{The detailed calculation procedure is shown in Appendix~\ref{B}.}
The error in one-dimensional reduced shear is given by
\begin{equation}
	\sigma_g\sim \frac{0.27\times 4\sqrt 2 ~\delta\alpha_s}{\sqrt N}\sim \frac{1.53 ~\delta\alpha_s}{\sqrt N}\,,
\end{equation}
where the angle $\delta\alpha_s$ is in radians. 
Comparing it with Eq.~(\ref{appendixa}), we conclude that the error in $\delta \alpha_s$ must be smaller than $1/4\sqrt 2$ radian ($\simeq 10^\circ$) for the new method to yield more accurate results. 
Note also that the degree of precision required for our new method to surpass the current method does not depend on the rms value of the one-dimensional reduced shear, as it is eliminated in the comparison.

\section{Discussion}\label{discussion}

Certainly, we do not know any convincing mechanism that would make the orientation of lensed galaxies within the vicinity of the Bullet Cluster on the celestial plane non-random.
However, even if they are indeed random, in case there are not enough galaxies in each small patch of sky where the average of reduced shear is calculated, it can lead to a substantial error in the estimation of gravitational lensing effects. 
Even if the average of $e^{(s)}_\alpha$ is zero, we certainly know that each value of $e^{(s)}_\alpha$ can be far from being zero.

As mentioned in Section~\ref{introduction}, our study is not the first one that suggests the improved weak lensing investigations based on the polarization data of radio waves from a galaxy. 
However, it is the first one that contains the formulas that can be directly applied in the same manner as done in the seminal paper~\cite{Clowe:2006eq} where it is claimed to prove the existence of dark matter based on the Bullet Cluster observation.

The authors of Refs.~\cite{Francfort:2021oog, Francfort:2022laa}, who made the suggestion based on the polarization data, expressed their results in terms of the shear and suggested measuring certain quantities and a method to obtain $\varepsilon$ from the image of a galaxy. 
However, the seminal Bullet Cluster paper~\cite{Clowe:2006eq} used the reduced shear instead of the shear, and used $Q_{ij}$, i.e., Eq.~(\ref{Qij}), instead of the method presented in Ref.~\cite{Francfort:2021oog} to characterize the ellipticity $\varepsilon$. 
It is worth to note that significant effort can be saved by employing our formula (\ref{2-esquareesquare}) as opposed to the one in the previous paper~\cite{Francfort:2021oog}. 
The data for $Q_{ij}$, from which one can obtain the right-hand side of Eq.~(\ref{2-esquareesquare}) by means of Eq.~(\ref{e1ie2}), are already available, as they were to the authors of Ref.~\cite{Clowe:2006eq}.
We presented our results only in terms of the reduced shear and $Q_{ij}$ as done in Ref.~\cite{Clowe:2006eq}.  
Note also that there is no method that allows to determine the convergence from the shear. 
As explained in Eq.~(\ref{kappa}), only the reduced shear can be used to determine the convergence, which then can be immediately converted to the surface mass density.

Furthermore, the substitution of shear with reduced shear, as demonstrated in this article, holds substantial significance, given that the difference between the two is far from negligible in the Bullet Cluster.
In Ref.~\cite{Clowe:2006eq}, contours of the constant convergence are drawn. 
The outermost contour has $\kappa=0.16$ and the innermost contour has $\kappa=0.37$. 
Considering that the reduced shear is the shear divided by $(1-\kappa)$, the differences between the reduced shear and the shear are as large as 16\% at the outer region and 37\% near one of the centers of the Bullet Cluster. 

The authors of Refs.~\cite{Brown:2010rr, Brown:2011db} also performed some work on gravitational lensing based on polarization, but they used a different kind of ``complex ellipticity'' with a different definition, which was not used in the analysis in Ref.~\cite{Clowe:2006eq}. 
Therefore, their work is not directly applicable to the reanalysis of the gravitational lensing effects of the Bullet Cluster in the manner of Ref.~\cite{Clowe:2006eq}.

In Ref.~\cite{Clowe:2006eq}, the data was analyzed in such a way that the availability of data at the time best allowed. 
No data about the original orientations of galaxies were available as the polarization directions of radio waves from them were not known. 
However, as forthcoming radio surveys will measure the polarization, the situation could be different. 
Actually, arcsecond-scale radio polarization observations are already possible~\cite{arcsecond}. 
So far, Ref.~\cite{Clowe:2006eq} has been regarded as one of the strongest pieces of evidence for dark matter. 
We anticipate that our research will establish a foundation for an enhanced analysis of gravitational lensing in cases like this.  
In particular, it will have great applications for the generic gravitational lensing analysis of clusters, not only limited to the Bullet Cluster.

\appendix
\section{Derivation of the relation between the reduced shear and $\Delta \alpha$}\label{appendixkeyformula}

By closely following the detailed derivation of Eq.~(22) in Ref.~\cite{Francfort:2021oog}, we present the derivation of Eq.~(\ref{g2g1}), with the shear replaced by the reduced shear. 
Eq.~(\ref{expgamma}) must be replaced and can be decomposed as
\begin{equation}
(1-\kappa)	\exp \left(\begin{array}{cc}
		g_1 & g_2\\
		g_2 & -g_1
	\end{array}\right)=(1-\kappa)\left(\begin{array}{cc}
		\cos\phi & -\sin\phi\\
		\sin\phi & \cos\phi
	\end{array}\right)\left(\begin{array}{cc}
		e^g & 0\\
		0 & e^{-g}
	\end{array}\right)\left(\begin{array}{cc}
		\cos\phi & \sin\phi\\
		-\sin\phi & \cos\phi
	\end{array}\right)\,.
\end{equation}
In the first order, we have 
\begin{equation}
g_1=g\cos(2\phi),\qquad g_2=g\sin(2\phi)\,.
\end{equation}
Then, the Jacobi map Eq.~(\ref{D}) can be written as
\begin{equation}
D=(1-\kappa) D_s R(-\psi-\phi) \exp(g \sigma_3)R(\phi)\,,
\label{Dappendix}
\end{equation}
where 
\begin{equation}
R(\chi)=\left(\begin{array}{cc}
		\cos\chi & \sin\chi\\
		-\sin\chi & \cos\chi
	\end{array}\right)\,, \qquad \sigma_3=\left(\begin{array}{cc}
		1 & 0\\
		0 & -1
	\end{array}\right)\,.
\end{equation}
Then, $\xi_s$ a vector on the screen space at the source and $\theta_o$ the observed angle have the relation \begin{equation}
    \xi_s=D\theta_o\,.
    \label{xis}
\end{equation}

Now, at the source position, the galaxy has eccentricity $\epsilon_s$. 
If we normalize the semi-minor axis to have length 1, the semi-major axis has length $(1-\epsilon_s^2)^{-1/2}$. 
Such an ellipse is given by the condition
\begin{equation}
\xi_s^T A_s \xi_s=1\,,
\label{xiAxi}
\end{equation}
where
\begin{equation}
A_s=R(-\alpha_s)\left(\begin{array}{cc}
		1-\epsilon_s^2 & 0\\
		0 & 1
	\end{array}\right) R(\alpha_s)\,.
\end{equation}
If we plug in Eq.~(\ref{xis}) into Eq.~(\ref{xiAxi}), we obtain
\begin{equation}
\theta_o^T A_o \theta_0=1,
\end{equation}
where
\begin{eqnarray}
A_o&=&D^T A_s D\,.
\end{eqnarray}
$A_o$ can be diagonalized as the following form:
\begin{equation}
A_o=NR(-\alpha_o)\left(\begin{array}{cc}
		1-\epsilon_o^2 & 0\\
		0 & 1
	\end{array}\right) R(\alpha_o)\,,
\end{equation}
where $N$ is defined as the larger of the two eigenvalues of $A_o$. 
In the first order in $\psi$ and $\gamma$, we obtain
\begin{equation}
\epsilon_o=\epsilon_s-g \cos[2(\alpha_s-\phi)]\frac{2(1-\epsilon_s^2)}{\epsilon_s}\,,\quad 
\alpha_o=\alpha_s-\psi+g \sin[2(\alpha_s-\phi)]\frac{2-\epsilon_s^2}{\epsilon_s^2}\,.
\end{equation}
From the above equation, we can obtain Eq.~(\ref{g2g1}) upon using the fact that $\epsilon$ (i.e., $\epsilon_o$) is approximately equal to $\epsilon_s$ for small $g$ and $\psi$ is negligible.

\section{Estimation of $\sigma_{g}$ for the new method}\label{B}

Let's estimate $\sigma_{g}$, the error of 1D $g$ for the new method. 
We closely follow Ref.~\cite{Brown:2010rr} where an error estimate for a mathematically similar but contently different case was considered. 
We will use the following notation:
\begin{equation}
	g\equiv\left(\begin{array}{c}
		g_1 \\ g_2
	\end{array}\right)\,, \qquad 
 \hat n_i\equiv\left(\begin{array}{c}\sin 2\alpha_s^i\\-\cos 2\alpha_s^i\end{array}\right)\,, \quad 
 n_i^{\parallel}=\left(\begin{array}{c}\cos 2\alpha_s^i\\\sin 2 \alpha_s^i\end{array}\right)\,.
\end{equation}
Then, Eq.~(\ref{g2g1}) can be re-expressed as
\begin{equation}
	-\hat n_i \cdot g = \sqrt{e_1^2+e_2^2}~\Delta \alpha^i\,,
 \label{nig}
\end{equation}
which is a linear regression. 
We have to find $g_1$ and $g_2$, which minimize the following quantity,
\begin{equation}
	\chi^2=\sum_i w_i [\hat n_i \cdot g +\sqrt{e_1^2+e_2^2}~\Delta\alpha^i]^2\,,
\end{equation}
where $w_i$ denotes the weight. 
Such $g$ is given by the following formula 
\begin{equation}
	g= A^{-1} b\,,
 \label{g}
\end{equation}
where
\begin{equation}
	A=\sum_i w_i \hat n_i \hat n_i^T\,, \qquad 
    b=\sum_i w_i \sqrt{e_1^2+e_2^2}~ (\alpha_s^i-\alpha_o^i) \hat n_i\,.
\end{equation}
If we vary Eq.~(\ref{g}), we get
\begin{equation}
	\delta g= A^{-1} \delta b+\delta (A^{-1}) b\,.
 \label{deltag}
\end{equation}
Here, we have
\begin{equation}
	\delta b=\sum_i w_i \delta\sqrt{e_1^2+e_2^2}~(-\Delta\alpha^i)\hat n_i + \sum_i w_i \sqrt{e_1^2+e_2^2}~ \delta\alpha_s^i \hat n_i+\sum_i w_i \sqrt{e_1^2+e_2^2}~(-\Delta \alpha^i)2\delta \alpha_s^i n_i^{\parallel}\,.\label{deltab}
\end{equation}
On the right-hand side of Eq.~(\ref{deltab}), the first term and the third term are on the order of $\delta\alpha_s^i \Delta \alpha^i$, while the second term is on the order of $\delta\alpha_s^i$. 
Thus, we will retain solely the second term. 
On the right-hand side of Eq.~(\ref{deltag}), the second term is on the order of $\delta\alpha_s^i \Delta \alpha^i$, while the first term is on the order of $\delta\alpha_s^i$. 
Thus, we shall retrain exclusively the first term. 

Putting all together, we get
\begin{equation}
	\delta g\approx\sum_i w_i \sqrt{e_1^2+e_2^2} ~\delta \alpha_s^i m_i\,,\qquad m_i\equiv A^{-1}\hat n_i\,.
\end{equation}
The weights satisfy $\sum_i w_i=1$. 
For a rough estimate, we can put $w_i=1/N$ where $N$ is the number of galaxies per patch. 
Thus,
\begin{equation}
	\delta g\approx \frac 1N\sum_i \sqrt{e_1^2+e_2^2} ~\delta \alpha_s^i m_i\,.
\end{equation}
Using the fact that $\delta \alpha_s^i$ and $m_i$ are independent, and the fact that $\langle \delta\alpha_s^i\rangle=\langle m_i\rangle=0$ and $\sigma_m=2$, we obtain
\begin{equation}
	\sigma_g=\frac{2}{\sqrt N}\left\langle e_1^2+e_2^2 \right\rangle^{1/2} \delta\alpha_s\,.\label{sigmagnew}
\end{equation}  
In Section~\ref{currentmethod}, we mentioned that the rms 1D reduced shear was around 0.27. 
As the two eigenvalues of the shear susceptibility tensor $P^{\gamma}_{\alpha\beta}=2(\delta_{\alpha\beta}-e_\alpha e_\beta)$ are 2 and approximately 2, we have $e_1\sim 2 g_1$ and $e_2\sim 2 g_2$.
Therefore,
\begin{equation}
	\left\langle e_1^2+ e_2^2 \right\rangle^{1/2} \sim (0.27\cdot 2)\sqrt{2}\,,
\end{equation}
which implies
\begin{equation}
	\sigma_g\sim \frac{0.27\times 4\sqrt 2 ~\delta\alpha_s}{\sqrt N}\sim \frac{1.53 ~\delta\alpha_s}{\sqrt N}\,.
\end{equation}

\acknowledgments
We thank Douglas Clowe, J\'er\'emie Francfort, Ruth Durrer, and Han-Gil Choi for helpful discussions. The work is supported by the National Research Foundation of Korea (NRF) [NRF-2019R1C1C1005073 and NRF-2021R1A4A2001897 (YY, JCP), NRF-2021R1A2C1094577 (HSH)] and by IBS under the project code, IBS-R018-D1 (YY, JCP).



\begin{thebibliography}{99}

	\bibitem{Clowe:2006eq}
	D.~Clowe, M.~Bradac, A.~H.~Gonzalez, M.~Markevitch, S.~W.~Randall, C.~Jones and D.~Zaritsky,
	``A direct empirical proof of the existence of dark matter,''
	Astrophys. J. Lett. \textbf{648} (2006), L109-L113
	doi:10.1086/508162
	[arXiv:astro-ph/0608407 [astro-ph]].

	
	
	\bibitem{Stil:2008ew}
	J.~M.~Stil, M.~Krause, R.~Beck and A.~R.~Taylor,
	``The Integrated Polarization of Spiral Galaxy Disks,''
	Astrophys. J. \textbf{693} (2009), 1392-1403
	doi:10.1088/0004-637X/693/2/1392
	[arXiv:0810.2303 [astro-ph]].

	
	\bibitem{Brown:2010rr}
	M.~L.~Brown and R.~A.~Battye,
	``Polarization as an indicator of intrinsic alignment in radio weak lensing,''
	Mon. Not. Roy. Astron. Soc. \textbf{410}, 2057 (2011)
	doi:10.1111/j.1365-2966.2010.17583.x
	[arXiv:1005.1926 [astro-ph.CO]].

	
	\bibitem{Francfort:2021oog}
	J.~Francfort, G.~Cusin and R.~Durrer,
	``Image rotation from lensing,''
	Class. Quant. Grav. \textbf{38} (2021) no.24, 245008
	doi:10.1088/1361-6382/ac33ba
	[arXiv:2106.08631 [astro-ph.GA]].

	
	
	\bibitem{Brown:2011db}
	M.~L.~Brown and R.~A.~Battye,
	``Mapping the dark matter with polarized radio surveys,''
	Astrophys. J. Lett. \textbf{735}, L23 (2011)
	doi:10.1088/2041-8205/735/1/L23
	[arXiv:1101.5157 [astro-ph.CO]].




\bibitem{SKA:2018ckk}
D.~J.~Bacon \textit{et al.} [SKA],
``Cosmology with Phase 1 of the Square Kilometre Array: Red Book 2018: Technical specifications and performance forecasts,''
Publ. Astron. Soc. Austral. \textbf{37}, e007 (2020)
doi:10.1017/pasa.2019.51
[arXiv:1811.02743 [astro-ph.CO]].


 \bibitem{Brown:2015ucq}
M.~L.~Brown, D.~J.~Bacon, S.~Camera, I.~Harrison, B.~Joachimi, R.~B.~Metcalf, A.~Pourtsidou, K.~Takahashi, J.~A.~Zuntz and F.~B.~Abdalla, \textit{et al.}
``Weak gravitational lensing with the Square Kilometre Array,''
PoS \textbf{AASKA14}, 023 (2015)
doi:10.22323/1.215.0023
[arXiv:1501.03828 [astro-ph.CO]].



\bibitem{SKAMagnetismScienceWorkingGroup:2020xim}
G.~Heald \textit{et al.} [SKA Magnetism Science Working Group],
``Magnetism Science with the Square Kilometre Array,''
Galaxies \textbf{8}, no.3, 53 (2020)
doi:10.3390/galaxies8030053
[arXiv:2006.03172 [astro-ph.GA]].


\bibitem{Perlick:2004tq}
	V.~Perlick,
	``Gravitational lensing from a spacetime perspective,''
	Living Rev. Rel. \textbf{7}, 9 (2004)
	[arXiv:1010.3416 [gr-qc]].

	
	
\bibitem{Kaiser:1994jm}
	N.~Kaiser,
	``Nonlinear cluster lens reconstruction,''
	Astrophys. J. Lett. \textbf{439}, L1 (1995)
	doi:10.1086/187730
	[arXiv:astro-ph/9408092 [astro-ph]].

	
\bibitem{Schneider:1994jj}
	P.~Schneider,
	``Cluster lens reconstruction using only observed local data,''
	Astron. Astrophys. \textbf{302}, 639 (1995)
	[arXiv:astro-ph/9409063 [astro-ph]].

	
\bibitem{Blandford:1991edc}
	R.~D.~Blandford, A.~B.~Saust, T.~G.~Brainerd and J.~V.~Villumsen,
	``The distortion of distant galaxy images by large-scale structure,''
	Mon. Not. Roy. Astron. Soc. \textbf{251}, no.4, 600-627 (1991)
	doi:10.1093/mnras/251.4.600
 
	
\bibitem{Bartelmann:1999yn}
	M.~Bartelmann and P.~Schneider,
	``Weak gravitational lensing,''
	Phys. Rept. \textbf{340}, 291-472 (2001)
	doi:10.1016/S0370-1573(00)00082-X
	[arXiv:astro-ph/9912508 [astro-ph]].

	
	
\bibitem{Kaiser:1994jb}
	N.~Kaiser, G.~Squires and T.~J.~Broadhurst,
	``A Method for weak lensing observations,''
	Astrophys. J. \textbf{449}, 460-475 (1995)
	doi:10.1086/176071
	[arXiv:astro-ph/9411005 [astro-ph]].

	
\bibitem{Clowe:2005zg}
	D.~Clowe, P.~Schneider, A.~Aragon-Salamanca, M.~Bremer, G.~De Lucia, C.~Halliday, P.~Jablonka, B.~Milvang-Jensen, R.~Pello and B.~Poggianti, \textit{et al.}
	``Weak lensing mass reconstructions of the eso distant cluster survey,''
	Astron. Astrophys. \textbf{451}, 395 (2006)
	doi:10.1051/0004-6361:20041787
	[arXiv:astro-ph/0511746 [astro-ph]].


 \bibitem{Hou}
S.~Hou, X.~L.~Fan and Z.~H.~Zhu,
``Gravitational Lensing of Gravitational Waves: Rotation of Polarization Plane,''
Phys. Rev. D \textbf{100}, no.6, 064028 (2019)
doi:10.1103/PhysRevD.100.064028
[arXiv:1907.07486 [gr-qc]].

    \bibitem{Schneider}
	P.~Schneider,
``Gravitational lensing as a probe of structure,''
[arXiv:astro-ph/0306465 [astro-ph]].



\bibitem{Francfort:2022laa}
J.~Francfort, G.~Cusin and R.~Durrer,
``A new observable for cosmic shear,''
JCAP \textbf{09}, 003 (2022)
doi:10.1088/1475-7516/2022/09/003
[arXiv:2203.13634 [astro-ph.CO]].
 
	
\bibitem{arcsecond}
	C. ~Stanghellini, D. ~Dallacasa, M. ~Bondi and R.~Della Ceca,
	``Arcsecond scale radio polarization of BL Lacertae objects''
	Astron. Astrophys. \textbf{325}, 911 (1997)
	





\end{thebibliography}
\end{document}